\documentclass[a4paper,11pt]{article}

\usepackage{natbib}

\usepackage{amssymb,type1cm}

\usepackage[dvipdfm,bookmarks=true,bookmarksopen=true]{hyperref}

\title{Computability of simple games: A complete investigation of the sixty-four possibilities\thanks{
Journal of Mathematical Economics (2011)
\href{http://dx.doi.org/10.1016/j.jmateco.2010.12.003}{doi:10.1016/j.jmateco.2010.12.003}
}}

\author{Masahiro Kumabe \\
Faculty of Liberal Arts, The Open University of Japan\\
2-11 Wakaba, Mihama-ku, Chiba City, 261-8586 Japan
\and 
H.  Reiju Mihara\thanks{Corresponding author.
The mail address is available on \href{http://www5.atwiki.jp/reiju/}{his site}.   \protect\\
\emph{URL:} \url{http://econpapers.repec.org/RAS/pmi193.htm} (H.R. Mihara).}\\
Kagawa University Library\\
Takamatsu 760-8525, Japan}

\date{February 2011}

\newcommand{\N}{\mathbb{N}}
\newcommand{\REC}{\mathrm{REC}}
\newcommand{\CRec}{\mathrm{CRec}}

\newcommand{\qed}{\enspace\enspace \vrule height 6pt width5pt
depth2pt}

\usepackage{theorem}

\newtheorem{prop}{Proposition}

\newtheorem{lemma}{Lemma}

{\theorembodyfont{\normalfont}
\newtheorem{definition}{Definition}
}

\newenvironment{proof}{\emph{Proof}.}{\qed\bigskip}

\begin{document} 

\maketitle 

\begin{abstract} 
Classify simple games into sixteen ``types'' in terms of the four conventional axioms:
monotonicity, properness, strongness, and nonweakness.
Further classify them into sixty-four classes in terms of 
finiteness (existence of a finite carrier) and algorithmic computability.
For each such class, we either show that it is empty or give an example of a game belonging to it.
We observe that if a type contains an infinite game, 
then it contains both computable ones and noncomputable ones.
This strongly suggests that computability is logically,
as well as conceptually, unrelated to the conventional axioms.

\emph{Journal of Economic Literature} Classifications:  C71, C69, D71, D90.

\end{abstract}

\pagebreak

\section{Introduction}

Shortly after proposing four ``independent'' axioms characterizing simple majority rule \citep{may52},
\citet{may53} made a complete investigation of the axioms.  
By a ``\emph{complete investigation} of the four axioms,''
we mean an investigation of all the sixteen ($2^4$) classes (of rules), formed by classifying
all the rules in terms of whether they satisfy each axiom.\footnote{\label{weak-indep}%
Despite Arrow's endorsement \citep[footnote~27, page~102]{arrow63},
complete investigations of a set of axioms are rare in the literature, such as social choice,
that adopts the axiomatic method.
It is common to say that an axiom (called A1) is ``independent'' of some other axioms if there
are (i) a rule satisfying A1 and the others and 
(ii) a rule violating A1 but satisfying the others \citep[Section~4.1.3]{thomson01}.} %
In particular, May showed that the four axioms are ``completely independent'' in the sense
that  each of the sixteen classes is nonempty.

In this paper, we provide a \emph{complete investigation} of six axioms for simple games.
A \emph{(simple) game}\footnote{Sometimes referred to as a ``voting game'' or
a ``simple coalitional game'' in the literature.} 
is a coalitional game that assigns either 1 or~0 to each coalition---those assigned~1 are winning coalitions
and those assigned~0 are losing coalitions.
Among the six axioms, four are conventional: 
\emph{monotonicity},  \emph{properness}, \emph{strongness}, and \emph{nonweakness}.
These axioms classify games into sixteen ($2^4$) classes, which we call \emph{(conventional) types}.
The other two are \emph{finiteness} (existence of a finite carrier)
and  \emph{computability}, which is the focus of this paper.
The results of the investigation (of all the $2^4\times 2^2=64$ classes) 
are summarized in Table~\ref{theresults} in Section~\ref{overview}.\footnote{%
\citet{kumabe-m08scw} continue the complete investigation, considering only computable games.
That paper asks which ``degrees of rationality'' are achievable in each of the thirty-two classes,
while the present paper asks whether each class is empty.}

To present what we can observe from Table~\ref{theresults},
we define what we mean by an axiom (namely computability) being independent of others
 (namely the four conventional axioms):
We say that ``computability is \emph{independent of} the four axioms
 (\emph{within} a class of games)'' if for each of the sixteen types,
there is a computable game of that type (in that class) if and only if 
there is a noncomputable game of that type (in that class).\footnote{%
This notion of independence generally requires examination of many more cases 
than that in footnote~\ref{weak-indep} (which examines just two cases).
Note that ``complete independence'' in May's sense of the six axioms cannot be achieved, 
since the four conventional axioms are not ``completely independent.''  
For example, it is well known that there exist no weak, nonproper games.}
Put differently, if computability is \emph{not} independent of the four axioms within a certain class, 
then for some type~$t$, there are type $t$ games in the class, but they are all computable 
or all noncomputable.

One of our main findings is (Proposition~\ref{prop:indep}) that 
\emph{computability is independent of the four conventional axioms within the class of \emph{infinite} games}.
(The analogue of Proposition~\ref{prop:indep} does not hold for the class of \emph{finite} games.  
This is because all finite games are computable.)
In fact, we come close to saying that computability is independent of the four conventional axioms
(within the class of \emph{all} games).  The conditions for the independence are satisfied for fifteen out of 
the sixteen types.  The only exception is type~2, consisting exclusively of dictatorial (hence computable) games.
This strongly suggests that computability is logically, as well as conceptually, 
unrelated to the conventional axioms.\footnote{%
What is behind this terminology is the discussion of logical and conceptual independence by \citet{thomson01}.
We do not define ``conceptual independence'' mathematically.}
In other words, as far as compatibility with the conventional axioms are concerned, 
computability is almost nonrestrictive.

\bigskip

The rest of the Introduction gives a brief background.
The companion paper~\citep{kumabe-m08jme} gives further discussion.

One can think of simple games as representing voting methods or
multi-criterion decision rules.
They have been central to the study of social choice \citep[e.g.,][]{peleg02hbscw,kumabe-m1008geb}.
For this reason, the paper can be viewed as a contribution to the foundations of 
\emph{computability analysis of social choice}, 
which studies algorithmic properties of social decision-making.\footnote{%
This literature includes \citet{kelly88}, \citet{lewis88}, \citet{bartholdi-tt89vs,bartholdi-tt89cd},
\citet{mihara97et,mihara99jme,mihara04mss}, and \citet{kumabe-m08jme,kumabe-m08scw}.}  %

The importance of computability in social choice theory would be unarguable.
First, the use of the language by social choice theorists suggests the importance: 
for example, \citet{arrow63} uses words such as ``\emph{process or rule}'' or
``\emph{procedure}.'' 
Second, there is a normative reason:
computability of social choice rules formalizes the notion of ``due process.''

We consider an infinite set of ``players.''
Roughly speaking, a simple game is \emph{computable} if there is a Turing program (finite algorithm) that can decide
from any description (by integer) of each coalition whether it is winning or losing.
Since each member of a coalition should be describable, we assume that the
set $N$ of (the names of) players is countable, say, $N=\N=\{0,1,2, \dots \}$.
Each coalition is described by a Turing program that can decide for the name of each player whether
she is in the coalition.  Note that there are infinitely many Turing programs that describes the same coalition.
Since each Turing program has its code number (G\"{o}del number),
the coalitions describable in this manner are describable by an integer, as desired.
(Such coalitions are called \emph{recursive} coalitions.)

\citet{kumabe-m08jme} give three interpretations of \emph{countably many players}:
(i)~generations of people extending into the indefinite future,
(ii) finitely many \emph{persons} facing countably many \emph{states} of the world~\citep{mihara97et}, and
(iii)~attributes or \emph{criteria} in multi-criterion decision-making.

Examples of multi-criterion decisions include (a)~forming a team to perform a particular task 
\citep{kumabe-m08jme},\footnote{%
This example illustrates that the desirability of the (conventional) axioms depends on the context.
Monotonicity makes sense here, but may be too optimistic (adding a member may turn an acceptable team into an
unacceptable one).  Properness may be irrelevant or even undesirable 
(ensuring that a given task can be performed by 
two non-overlapping teams may be important from the viewpoint of reliability).
These observations suggest the importance of finding games that violate some of the axioms.}  %
(b)~granting tenure to junior faculty members at academic institutions~\citep{alnajjar-af06},
and 
(c)~deciding whether a certain act is legal \citep{kumabe-m07csg64}.
In these examples, there are potentially infinitely many criteria or contingencies on which decisions can be based.

\section{Framework} 

\subsection{Simple games}\label{notions}

Let $N=\N=\{0,1,2, \dots \}$ be a countable set of (the names of) 
players.
Intuitively, a simple game describes in a crude manner the power 
distribution among observable (or describable) subsets of players.
Such sets are called \emph{coalitions}.
In this paper, we define a \textbf{coalition} to be a recursive (algorithmically decidable) set; it is 
a set of players for which there is a Turing program (algorithm)
that can decide for the name of each player whether she is in the set.\footnote{%
A set~$S$ is \emph{recursive} if there is a Turing machine that halts on any input $i\in N$,
yielding output 1 if $i\in S$ and 0 otherwise.
\citet{soare87} and \citet{odifreddi92}
give a precise definition of recursive sets as well as detailed discussion of recursion theory.
Mihara's papers~\citep{mihara97et,mihara99jme} contain short reviews of recursion theory.} 
Note that \textbf{the class~$\REC$ of (recursive) coalitions} 
forms a \textbf{Boolean algebra}; that is, it includes $N$ 
and is closed under union, intersection, and complementation.

Formally, a \textbf{(simple) game} is a collection~$\omega\subseteq\REC$ of (recursive) coalitions.
We will be explicit when we require that $N\in \omega$.
The coalitions in $\omega$ are said to be \textbf{winning}.  
The coalitions not in $\omega$ are said to be \textbf{losing}. 
One can regard a simple game as a function from~REC to $\{0,1\}$, assigning the value 1 or 0 to each 
coalition, depending on whether it is winning or losing.

We introduce from the theory of cooperative games a few basic 
notions of simple games~\citep{peleg02hbscw,weber94}.
A simple game $\omega$ is said to be 
\textbf{monotonic} if for all coalitions $S$ and $T$, the 
conditions $S\in \omega$ and $T\supseteq S$ imply $T\in\omega$.  
$\omega$ is \textbf{proper} if for all recursive coalitions~$S$, 
$S\in\omega$ implies $S^c:=N\setminus S\notin\omega$.  $\omega$ is 
\textbf{strong} if for all coalitions~$S$, $S\notin\omega$ 
implies $S^c\in\omega$.  $\omega$ is \textbf{weak} if 
$\omega=\emptyset$ or
the intersection~$\bigcap_{S\in\omega}S=\bigcap\omega$ of the winning coalitions is nonempty.  
The members of $\bigcap_{S\in\omega}S$ are called \textbf{veto players}; they 
are the players that belong to all winning coalitions.  
(The set $\bigcap_{S\in\omega}S$ of veto players may or may not be observable.)
$\omega$ is \textbf{dictatorial} if there exists some~$i_0$ (called a 
\textbf{dictator}) in~$N$ such that $\omega=\{\,S\in\REC: i_0\in 
S\,\}$.  Note that a dictator is a veto player, but a veto player is 
not necessarily a dictator.
It is immediate to prove the following well-known lemmas:

\begin{lemma} \label{weakisproper}
If a simple game is weak, it is proper.\end{lemma}

\begin{lemma} \label{strongweakisdic}
A simple game is dictatorial if and only if it is strong and weak.\end{lemma}

A \textbf{carrier} of a simple game~$\omega$ is a coalition $S\subseteq N$ 
such that for all coalitions~$T$, we have $T\in\omega$ iff $S\cap T\in \omega$.
We observe that if $S$ is a carrier, then so is any coalition $S'\supseteq S$.
Slightly abusing the word, we sometimes say a game is \textbf{finite} if it has a finite carrier; 
otherwise, it is \textbf{infinite}.

\subsection{The computability notion}
\label{comp:notion}

\noindent\textbf{Notation}. 
A \emph{partial function (of $n$ variables)} is a function (into natural numbers) 
whose domain is a subset of~$\N^n$.
For a partial function $\psi$, $\psi(x)\downarrow$  means $\psi(x)$ is defined;
$\psi(x)\uparrow$ means $\psi(x)$ is undefined.
For $k\in \N$, let
$\varphi_k(\cdot)$ be the \emph{$k$th partial recursive function} (of one variable)---it is 
the partial function (of one variable) computed by the Turing program with code
(G\"{o}del) number~$k$.\enspace$\|$

\medskip

First, we represent each recursive coalition by a characteristic index ($\Delta_0$-index).  
A number $e$~is a \textbf{characteristic index} for a coalition~$S$
if $\varphi_e$
is the characteristic function for~$S$.\footnote{The \emph{characteristic function} for~$S$ takes the value 1
if the input belongs to~$S$; it takes 0 otherwise.
The same coalition has infinitely many characteristic indices.}
Intuitively, a characteristic index for a coalition describes
the coalition by a Turing program that can decide its membership.

Next, we introduce an indicator for a game.
It assigns the value 1 or 0 to each 
number representing a coalition, depending on whether the 
coalition is winning or losing.  When a number does not represent a 
recursive coalition, the value is undefined.
Given a simple game $\omega$, its \textbf{$\delta$-indicator} is the partial 
function~$\delta_\omega$ on~$\N$ defined by
\[
	\label{d:eq}
	\delta_\omega(e)=\left\{
	    \begin{array}{ll}
		1 & \mbox{if $e$ is a characteristic index for a recursive
	set in $\omega$},  \\
		0 & \mbox{if $e$ is a characteristic index for a recursive
	set not in $\omega$},  \\
		\uparrow & \mbox{if $e$ is not a characteristic
	index for any recursive set}.
	\end{array}
	\right.
\]
Note that $\delta_\omega$ is well-defined since each $e\in\N$ can be a 
characteristic index for at most one set.
If $e$ and $e'$ are characteristic indices for the same coalition, then the
definition implies $\delta_\omega(e)=\delta_\omega(e')$.

Finally, we introduce the notion of \emph{($\delta$)-computable} games.
The condition requires existence of a Turing program that correctly answers whether a coalition
is winning or losing, from any one of infinitely many characteristic indices for the coalition.

\begin{definition} A game $\omega$ is ($\delta$)-\textbf{computable} if  
$\delta_\omega$ has an extension to a partial recursive function.\footnote{%
A partial function $\delta'$ is an \emph{extension} of $\delta_\omega$ if whenever 
$\delta_\omega(e)\downarrow$, we have $\delta'(e)=\delta_\omega(e)$.}
\end{definition}

Among various notions of computability that we could conceive of, 
this notion is the only one that we find~\citep{mihara04mss} defensible.\footnote{
As long as games are defined for (recursive) coalitions, this notion of computability is
equivalent to the following \citep[Corollary~1]{kumabe-m07csg64}:
there exists a Turing machine that, given any coalition~$S$ encoded as an infinite binary sequence
($i$th term indicating whether $i\in S$),
halts and correctly decides whether $S$~is winning.}

\section{Overview of the Results}\label{overview}

This section gives a summary of the results in Sections~\ref{finitecarriers}--\ref{nofinitecarriers}.

We classify games into sixty-four ($2^4\times 2^2$) classes as shown in~Table~\ref{theresults},
in terms of their \textbf{(conventional) types} 
(with respect to the conventional axioms of monotonicity, properness, strongness, and nonweakness), 
\emph{finiteness} (existence of a finite carrier), and $\delta$-\emph{computability}.
For each of the 64 classes, we ask whether there exists a game in the class.
The answers are given in Sections \ref{finitecarriers}--\ref{nofinitecarriers}.\footnote{\label{emptytypes}
Among the sixteen types, five (types 6, 8, 10, 14, and 16) contain no games;
also, the class of type~$2$ infinite games is empty (since type 2 games are dictatorial).
These results are immediate from Lemmas \ref{weakisproper} and~\ref{strongweakisdic}.}
Table~\ref{theresults} summarizes the answers.\footnote{%
Some of the games constructed in this paper have the property that an empty coalition is winning.
However, one can modify all such computable games so that an empty coalition is losing \citep{kumabe-m08scw}.
}

\begin{table}
\caption{Existence of Games in Different Classes}
\begin{center}
\begin{tabular}{rccccc} \hline\hline
          & \multicolumn{2}{c}{Finite} & & \multicolumn{2}{c}{Infinite} \\
	\cline{2-3} \cline{5-6}
Types	& Non & Computable & & Non & Computable \\
 \hline
 1 $(++++)$ & no & yes & & yes & yes \\
 2 $(+++-)$ & no & \emph{yes} & & \emph{no} & \emph{no} \\
 3 $(++-+)$ & no & yes & & yes & yes \\
 4 $(++--)$ & no & yes & & yes & yes \\
 5 $(+-++)$ & no & yes & & yes & yes \\
 6 $(+-+-)$ & no & no & & no & no \\
 7 $(+--+)$ & no & yes & & yes & yes \\
 8 $(+---)$ & no & no & & no & no \\
 9 $(-+++)$ & no & yes & & yes & yes \\
 10 $(-++-)$ & no & no & & no & no \\
 11 $(-+-+)$ & no & yes & & yes & yes \\
 12 $(-+--)$ & no & yes & & yes & yes \\
 13 $(--++)$ & no & yes & & yes & yes \\
 14 $(--+-)$ & no & no & & no & no \\
 15 $(---+)$ & no & yes & & yes & yes \\
 16 $(----)$ & no & no & & no & no \\ \hline \\
\end{tabular}
\parbox{110mm}{\footnotesize
The types are defined by the four conventional axioms: 
monotonicity, properness, strongness, and nonweakness. 
For example, the entries corresponding to type~2 $(+++-)$
 indicates that among the monotonic ($+$), proper ($+$), strong ($+$), weak ($-$, because not nonweak) games, 
there exist no finite noncomputable ones,
there exist finite computable ones, 
there exist no infinite noncomputable ones, and
there exist no infinite computable ones.
Note that except for type~2, the last three columns are identical.}  %
\end{center}\label{theresults}
\end{table}

We are mainly interested in the relation of computability to the four conventional axioms.
What can we observe from Table~\ref{theresults}?  
For example, we can see that 
\emph{there is a computable game of type~2 $(+++-)$, but not a noncomputable game of the same type}.
(In fact, type~2 consists of dictatorial games.)
This means that computability is \emph{not} ``independent of'' the four axioms in the following sense:
there is a nonempty type consisting only of computable games or only of noncomputable games.

\emph{For each of the other fifteen types, however, 
there is a computable game of that type if and only if there is a noncomputable game of that type}.
Hence, we could almost say that computability is ``unrelated to'' the four axioms.
In fact, if we restrict our attention to the infinite games (games without a finite carrier), 
we can say this:

\begin{prop} \label{prop:indep}
The axiom $\delta$-computability is \emph{independent of} monotonicity, properness, 
strongness, and nonweakness within the class of infinite games 
in the following sense:  for each of the $2^4=16$ types,
there exists a computable infinite game of that type if and only if 
there exists a noncomputable infinite game of that type.\end{prop}

We leave this section with two interesting observations involving the last three
(instead of two as in Proposition~\ref{prop:indep}) columns of the table:
From the rows corresponding to types 6, 8, 10, 14, 16, we conclude that 
\emph{if there does not exist a finite computable game of a particular type, 
then there does not exist a game of that type}.
From the other rows except row~2, we conclude that
\emph{if there exists an infinite (non)computable game of a particular type, 
then there exists a finite computable game of that type}.

\section{Preliminary Results}
\label{prelim}

This section gives a sufficient condition and a necessary condition for a game to be computable.
It also introduces notation needed in Sections~\ref{finitecarriers}--\ref{nofinitecarriers}.  

\medskip

\noindent\textbf{Notation}.  We identify a natural number~$k$ with the finite 
set $\{0,1,2,\ldots,k-1\}$, which is an initial segment of~$\N$.  
Given a coalition $S\subseteq N$, we write $S\cap k$ to represent the 
coalition $\{i\in S: i<k\}$ consisting of the members of $S$ whose 
name is less than~$k$.  
We call $S\cap k$ the \textbf{$k$-initial segment of $S$}, and view it 
either as a subset of~$\N$ or as the string $S[k]$ of length~$k$ of 0's and 1's 
(representing the restriction of its characteristic function to 
$\{0,1,2,\ldots,k-1\}$).\enspace$\|$

\begin{definition}
Consider a simple game.  A string $\tau$ (of 0's and 1's) of 
length~$k\geq 0$ is \textbf{winning determining} if any 
coalition $G\in\REC$ extending $\tau$ (in the sense that $\tau$ is an 
initial segment of $G$, i.e., $G\cap k=\tau$) is winning; $\tau$ is 
 \textbf{losing determining} if any 
coalition $G\in\REC$ extending $\tau$ is losing.  
A string is \textbf{determining} if it is either winning determining or losing determining.\end{definition}

First, \emph{to construct computable games}, we use the following proposition,
which simply restates the ``if'' direction of Theorem~4 in \citet{kumabe-m08jme}.
In particular, \emph{finite games are computable}.
As seen in Section~\ref{overview}, whether a game is finite 
is an important criterion for classifying games in this paper.

\begin{prop} \label{delta0det2}
Let $T_0$ and $T_1$ be recursively enumerable sets of (nonempty) strings such that
any coalition has an initial segment in $T_0$ or in $T_1$ but not both.
Let $\omega$ be the simple game defined by $S\in \omega$ if and only if
 $S$ has an initial segment in~$T_1$.
Then $T_1$ consists only of winning determining strings, $T_0$ consists only of losing determining strings
(so $S\notin \omega$ if and only if $S$ has an initial segment in~$T_0$), 
and $\omega$ is $\delta$-computable.\end{prop}

Second, \emph{to construct noncomputable games}, we use the following proposition
 \citep[Proposition~3]{kumabe-m08jme}.
Here, the number~$k-1$ may 
be greater than the greatest element, if any, of~$S$:

\begin{prop} \label{cutprop}
Suppose that a $\delta$-computable simple game is given.  
\textup{(i)}~If a coalition $S$ is winning, then it has an initial segment $S[k]$ (for some $k\in \N$) 
that is winning determining.
\textup{(ii)}~If $S$ is losing, then it has an initial segment $S[k]$ that is losing determining.
\end{prop}

\medskip

\noindent\textbf{Notation}.  
Let $\alpha$ and $\beta$ be strings (of 0's and 1's).
Then $\alpha^c$ denotes the string of the length $|\alpha|$ such that $\alpha^c(i)=1-\alpha(i)$ for each $i<|\alpha|$; 
for example, $0110100100^c=1001011011$.
Occasionally, a string $\alpha$ is identified with the set $\{i: \alpha(i)=1\}$.
(Note however that $\alpha^c$ is occasionally identified with the set $\{i: \alpha(i)=0\}$,
but never with the set $\{i: \alpha(i)=1\}^c$.)
$\alpha\beta$ (or $\alpha*\beta$) denotes the concatenation of $\alpha$ followed by $\beta$.
$\alpha[k]$ denotes the prefix (initial segment) of $\alpha$ of length~$k$.
$\alpha\subseteq \beta$ means that $\alpha$ is a prefix of $\beta$ ($\beta$ extends $\alpha$);
$\alpha \subseteq A$, where $A$ is a set, means that $\alpha$ is an initial segment of~$A$
(i.e, $\alpha$ is equal to the initial segment $A[k]$, for some $k$.)
Strings $\alpha$ and $\beta$ are \textbf{incompatible} if neither
$\alpha\subseteq \beta$ nor $\beta\subseteq\alpha$
(i.e., there is $k< \min\{|\alpha|,|\beta|\}$ such that $\alpha(k)\neq \beta(k)$).\enspace$\|$

\section{Finite Games}
\label{finitecarriers}

We start with the class of finite games (games having a finite carrier).  
Any game in this class is $\delta$-computable.

In the following,
for each of the eleven conventional types (with respect to monotonicity, properness, strongness, and nonweakness)
not shown to be empty so far (footnote~\ref{emptytypes}),
we give an example of a finite game of that type 
by exhibiting finite sets~$T_0$ and $T_1$ satisfying the condition of Proposition~\ref{delta0det2}.

\begin{enumerate}
\item [1] $(++++)$
A monotonic, proper, strong,  nonweak game.  Let $T_0=\{00, 010, 100\}$ and $T_1=\{11, 011, 101\}$. 

\item [2] {$(+++-)$} A monotonic, proper, strong, weak game.  
Let $T_0=\{0\}$ and $T_1=\{1\}$.  Player~$0$ is a dictator.  

\item [3] {$(++-+)$} A monotonic, proper, nonstrong, nonweak game.
Let $T_0=\{00, 010, 0110, 100, 1010\}$ and $T_1=\{11, 1011, 0111\}$.  

\item [4] {$(++--)$} A monotonic, proper, nonstrong, weak game.
Let $T_0=\{0, 10\}$ and $T_1=\{11\}$.  

\item [5] {$(+-++)$} A monotonic, nonproper, strong, nonweak game.
Let $T_0=\{00\}$ and $T_1=\{1, 01\}$.  

\item[7] {$(+--+)$} A monotonic, nonproper, nonstrong, nonweak game.
Let $T_0=\{00,100,0110,0100\}$ and 
$T_1=\{11,101,0101,0111\}$. 

\item [9] {$(-+++)$} A nonmonotonic, proper, strong, nonweak game.
Let $T_0=\{1\}$ and $T_1=\{0\}$. 

\item [11] {$(-+-+)$} A nonmonotonic, proper, nonstrong, nonweak game.
Let $T_0=\{1, 01\}$ and $T_1=\{00\}$. 

\item [12] {$(-+--)$} A nonmonotonic, proper, nonstrong, weak game.
Let $T_0=\{1, 00\}$ and $T_1=\{01\}$.  

\item [13] {$(--++)$} A nonmonotonic, nonproper, strong, nonweak game.
Let $T_0=\{10\}$ and $T_1=\{0, 11\}$.  

\item [15] {$(---+)$} A nonmonotonic, nonproper, nonstrong, nonweak game.
Let $T_0=\{01, 10\}$ and $T_1=\{00, 11\}$.  

\end{enumerate}

\section{Infinite Games}\label{nofinitecarriers}

We consider infinite games (games without finite carriers) in this section.

\subsection{Noncomputable games} \label{infinitenoncomp}

We first give examples of infinite \emph{noncomputable} simple games.
Proposition~\ref{cutprop} implies that all \emph{computable} games 
(that have both winning and losing coalitions) belong to the class of games 
that have both finite winning coalitions and cofinite losing coalitions.
To show that variety is not lost even if we restrict our games to this class,
all the examples are chosen from the class.
The examples in this section are based on the following lemma.

\begin{lemma}\label{typicalnoncomp}
Let $A$ be a recursive set.
Let $T_0$ and $T_1$ be recursively enumerable, nonempty sets of (nonempty) strings such that
any coalition has an initial segment in $T_0$ or in $T_1$ but not both.
Let $\omega$ be the simple game defined by $S\in \omega$ if and only if either
$S=A$ or [$S \ne A^c$ and $S$ has an initial segment in~$T_1$].
Then we have the following:\\
\textup{(i)}~$S\notin\omega$ if and only if either 
$S=A^c$ or [$S \ne A$ and $S$ has an initial segment in~$T_0$].\footnote{%
Let $\hat{\omega}$ be the game defined by Proposition~\ref{delta0det2}.
It follows that (a)~$S\in \omega$ if and only if either
$S=A$ or [$S \ne A^c$ and $S\in \hat{\omega}$],
(b)~$S\notin\omega$ if and only if either 
$S=A^c$ or [$S \ne A$ and $S\notin\hat{\omega}$],
(c)~if $\hat{\omega}$ is proper, then $\omega$ is proper,
(d)~if $\hat{\omega}$ is strong, then $\omega$ is strong.}\\
\textup{(ii)}~$\omega$ has a finite winning coalition and a cofinite losing coalition.\\
\textup{(iii)}~Suppose further that either $A$ is infinite and has an initial segment in~$T_0$ or 
$A^c$ is infinite and has an initial segment in~$T_1$. Then $\omega$ is $\delta$-noncomputable (hence infinite).
\end{lemma}

\begin{proof} 
(i)~From the definition of $\omega$ 
and the assumption that any coalition $S$ has a initial segment in $T_0$ or $T_1$ but not both, we have
\begin{eqnarray*}
S\notin \omega & \iff & \textrm{$S\neq A$ and [$S=A^c$ or $S$ has no initial segment in~$T_1$]} \\
					& \iff & \textrm{[$S\neq A$ and $S=A^c$] or} \\
					& & 	\textrm{[$S\neq A$ and $S$ has no initial segment in~$T_1$]} \\
					& \iff & \textrm{[$S=A^c$] or [$S\neq A$ and $S$ has an initial segment in~$T_0$].}
\end{eqnarray*}

(ii)~Choose a string $\alpha$ from the nonempty set~$T_1$.
Let $\beta=\alpha*A(|\alpha|)$.
Then $\beta\ne A^c$ since $\beta(|\alpha|)=A(|\alpha|)\neq A^c(|\alpha|)$.
Since $\beta$ has the prefix (initial segment) $\alpha\in T_1$, 
 $\beta\in \omega$ by the definition of $\omega$.
We have obtained a finite winning coalition, namely~$\beta$.
To obtain a cofinite losing coalition, choose $\alpha\in T_0$ and let $\beta=\alpha*A^c(|\alpha|)$.
Then by~(i), $B:=\{i: \textrm{$\beta(i)=1$ or $\beta(i)\uparrow$} \}$ is a cofinite losing set.

(iii)~Suppose $A$ is infinite and has an initial segment $A[k]$ in~$T_0$.
Suppose $\omega$ is $\delta$-computable.  Then, by Proposition~\ref{cutprop},
the winning coalition $A$ has an initial segment $A[k']$ that is a winning determining string.
Let $\hat{k}=\max\{k, k'\}$.  
Then on the one hand, $A[\hat{k}]$, 
which is different from $A$ and has an initial segment in $T_0$, is losing by~(i).  
On the other hand, $A[\hat{k}]$ is winning since it extends the winning determining string~$A[k']$.
We have obtained a contradiction.
The case where $A^c$ is infinite and has an initial segment in~$T_1$ is similar.\end{proof} %

For each conventional type~$t$ not shown to be empty so far (there are ten such types; footnote~\ref{emptytypes}),
we can construct an example of an infinite noncomputable game $\omega^t$ of that type as follows:
Let $T_0$ and $T_1$ be those sets in the example for type~$t$ in Section~\ref{finitecarriers}.
Let $A$ be the infinite set represented by $\tau*1111\ldots$ (i.e., $i\notin A$ iff $i<|\tau|$ and $\tau(i)=0$),
where $\tau$ is any string belonging to~$T_0$.
(For $t=7$, we also require $\tau\neq 0100$.)
For $t\ne 5$, let $\omega^t$ be the game $\omega$ defined by Lemma~\ref{typicalnoncomp}.
For $t=5$, define $\omega^5$ by $S\in \omega^5$ if and only if $S=A$ or $S$ has an initial segment in~$T_1$
(thus $S\notin\omega^5$ if and only if $S\ne A$ and $S$ has an initial segment in~$T_0$).
It is routine to verify, for each $t$, that $\omega^t$ is indeed of type~$t$.\footnote{\citet{kumabe-m07csg64} 
give more detailed proofs for a different set of examples.}

\subsection{A class of infinite, computable, type~1 games}
\label{nice_games}

In this section, \emph{we construct for each recursive set~$A$, 
an infinite, computable, monotonic, proper, strong, nonweak simple game $\omega[A]$}. 
The construction is self-contained, but long and elaborate.
One reason that the construction is complicated is that we construct a \emph{family} of type~1 games $\omega[A]$, 
one for each recursive set~$A$,
while requiring \emph{additional conditions} that would become useful for constructing other types of
games in Section~\ref{examples:nocarrier}.\footnote{%
In \citet[Appendix~A]{kumabe-m08scw}, we construct just one type~1 game, 
without requiring the additional conditions.
Some aspects of the construction thus become more apparent in that construction.
The construction there extends the one 
(not requiring the game to be of a particular type) in the companion 
paper~\citep[Section~6.2]{kumabe-m08jme}.
The reader might want to consult these papers first.} %

\bigskip

Our approach is to construct recursively enumerable sets $T_0$ and $T_1$ 
of strings (of 0's and 1's) satisfying the conditions of Proposition~\ref{delta0det2}.
We first construct certain sets~$F_s$ of strings for $s\in\{0, 1, 2, \ldots\}$.
We then specify an algorithm for enumerating the elements of $T_0$ and $T_1$ using the sets~$F_s$, 
and construct a simple game $\omega[A]$ according to Proposition~\ref{delta0det2}.
We conclude that the game is computable by checking (Lemma~\ref{k46}) 
that $T_0$ and $T_1$ satisfy the conditions of Proposition~\ref{delta0det2}.
Finally, we show (Lemmas~\ref{k47a}, \ref{k47b}, and~\ref{k48}) 
that the game satisfies the desired properties.

Before constructing sets $T_0$ and $T_1$ of determining strings, we introduce the notions of p-strings and d-strings.
Roughly speaking, a p-string consists of $10$'s or $01$'s; 
A d-string is a concatenation of a p-string followed by $00$ or~$11$.
More formally, a string $\alpha$ is a \textbf{p-string} if $|\alpha|$ is even and for each $2k<|\alpha|$, we have
$\alpha(2k) \alpha(2k+1)\in \{10, 01\}$ (i.e., $\alpha(2k+1)=1-\alpha(2k)$).  
Examples of a p-string include the empty string, 01, 0101, 0110, and 1001011010.  
Note that any prefix (initial substring) of even length of a p-string is a p-string. 
Denote by $\alpha^-$ the prefix $\alpha[|\alpha|-1]$ of $\alpha$ of length $|\alpha|-1$.  In other words, 
$\alpha=\alpha^-*\alpha(|\alpha|-1)$.
A string $\alpha$ (of even length) is a \textbf{d-string} if $\alpha^{--}$ is a p-string
and $\alpha(|\alpha|-2) \alpha(|\alpha|-1)\in \{00, 11\}$ (i.e., $\alpha(|\alpha|-2)=\alpha(|\alpha|-1)$).
In other words,  a d-string $\alpha$ is of the form $\alpha^{--}*00$ or $\alpha^{--}*11$
for some p-string~$\alpha^{--}$.
It is easy to prove \citep{kumabe-m07csg64} the following lemma:

\begin{lemma}\label{k41}
\textup{(i)}~Any string of even length either is a p-string or extends a d-string.
\textup{(ii)}~Any two distinct d-strings $\alpha$ and $\beta$ are incompatible.
That is, we have neither $\alpha\subseteq \beta$ nor $\beta\subseteq\alpha$
(i.e., there is $k< \min\{|\alpha|,|\beta|\}$ such that $\alpha(k)\neq \beta(k)$).
 \end{lemma}

Let $\{k_s\}_{s=0}^\infty$ be an effective listing (recursive enumeration) of the members of 
 the recursively enumerable set $\{k : \varphi_k(2k)\in \{0,1\} \}$,
 where $\varphi_k(\cdot)$ is the $k$th partial recursive function of one variable (which is computed 
 by the Turing program with code number~$k$).
We can assume without loss of generality that $k_0\geq 1$ and all the elements $k_s$ are distinct.
Thus, 
\[ \CRec \subset \{k : \varphi_k(2k)\in \{0,1\}\} = \{k_0, k_1, k_2, \ldots\}, \]
where $\CRec$ is the set of characteristic indices for recursive sets.

Let $l_{0}=2k_0+2\geq 4$ and for $s>0$, let $l_{s}=\max\{l_{s-1}, 2k_{s}+2\}$. 
Then $\{l_s\}$ is an nondecreasing sequence of even numbers and
$l_s>2k_s+1$ for each~$s$.  Note also that $l_s\geq l_{s-1}>2k_{s-1}+1$, $l_s\geq l_{s-2}>2k_{s-2}+1$, etc.\ 
imply that $l_s> 2k_s+1$, $2k_{s-1}+1$, $2k_{s-2}+1$, \ldots, $2k_0+1$.

For each~$s$, let $F_s$ be the finite set of  p-strings 
$\alpha=\alpha(0)\alpha(1)\cdots\alpha(l_s-1)
\supseteq 10$ of length $l_{s}\ge 4$ such that 
\begin{enumerate}
\item [(1)] $\alpha(2k_s)=\varphi_{k_s}(2k_s)$ and for each $s'<s$,  $\alpha(2k_{s'})=1-\varphi_{k_{s'}}(2k_{s'})$.
\end{enumerate}
Note that (1) imposes no constraints on $\alpha(2k)$ for $k\notin\{k_0,k_1,k_2, \ldots, k_s\}$, 
while it actually imposes constraints for all $k$ in the set, 
since $|\alpha|=l_s> 2k_s$, $2k_{s-1}$, $2k_{s-2}$, \ldots, $2k_0$.
We observe that if $\alpha\in F_s\cap F_{s'}$, then $s=s'$.
Let $F=\bigcup_{s}F_s$. Then $F$ is recursive and we have the following:

\begin{lemma} \label{Fincompatible1}
Any two distinct elements in $F$ are incompatible.
\end{lemma}

\begin{proof} 
Let $\alpha$, $\beta\in F$ such that $|\alpha|\leq |\beta|$, without loss of generality.  
 If $\alpha$ and $\beta$ have the same length, then the 
 conclusion follows since otherwise they become identical strings.
 If $l_s=|\alpha|< |\beta|=l_{s'}$, then $s<s'$ and by (1),
 $\alpha(2k_s)=\varphi_{k_s}(2k_s)$ on the one hand, but 
$\beta(2k_s)=1-\varphi_{k_s}(2k_s)$ on the other hand.  So $\alpha(2k_s)\neq \beta(2k_s)$.\end{proof}

Let $f$ be a recursive bijection from $F$ onto $\N$ 
($f$ can be obtained by enumerating the elements of $F$ one by one,
assigning $0$ to the first element enumerated, $1$ to the second element enumerated, and so on). 
Regarding $f$ as a partial function on the set of strings, 
we have $f(\alpha)\downarrow$ (i.e., $f(\alpha)$ is defined) if and only if $\alpha\in F$.

\begin{lemma} \label{k42}
Let $\alpha\supseteq 10$ be a p-string of length $l_s$. Then the following statements are equivalent:
\textup{(i)}~no prefix of $\alpha$ is in $F$;
\textup{(ii)}~for each $s'\leq s$, $\alpha[l_{s'}]\notin F$;
\textup{(iii)}~for each $s'\leq s$, $f(\alpha[l_{s'}])\uparrow$;
\textup{(iv)}~for each $s'\leq s$,  $\alpha(2k_{s'})=1-\varphi_{k_{s'}}(2k_{s'})$.
 \end{lemma}

\begin{proof}
The definition of $F$ implies that $\alpha\in F$ only if $|\alpha|=l_s$ for some $s$.  
Hence the equivalence of (i), (ii), and (iii) is immediate.
We next show that (ii) and (iv) are equivalent.  
The direction from (iv) to (ii) is clear from~(1).  
To see the other direction, suppose that (iv) is not the case; we derive the negation of~(ii).
For some $s'\leq s$, we have $\alpha(2k_{s'})=\varphi_{k_{s'}}(2k_{s'})$.  Choose the least such $s'$.
Then ($s'=0$ or) for any $s''<s'$, $\alpha(2k_{s''})=1-\varphi_{k_{s''}}(2k_{s''})$. 
So $\alpha[l_{s'}]\in F_{s'}$ by~(1), since $\alpha[l_{s'}]\supseteq 10$ is a p-string of length~$l_{s'}$. 
Thus (ii) is violated.\end{proof}

Let $A$ be a recursive set.
The game $\omega[A]$ will be defined via the sets $T_0:=T_0^A$ and $T_1:=T_1^A$
of strings, constructed by enumerating the elements as follows:

\textbf{Construction of $T_0$ and $T_1$}.
For each~$s$ and $\alpha\in F_s$ (having a length~$l_s$ and extending $10$), 
\begin{enumerate}
\item [(2.i)] for each p-string $\alpha'$ that is a proper prefix of $\alpha$, 
	if $s=0$ or $|\alpha'|\geq l_{s-1}$, then enumerate $\alpha'*11$ in $T_1$ and $\alpha'*00$ in~$T_0$;
\item [(2.ii)] if $f(\alpha)\in A$, enumerate $\alpha$ in $T_1$; 
if $f(\alpha)\notin A$, enumerate $\alpha$ in $T_0$ (note that $f(\alpha)\downarrow$ since $\alpha\in F$);
\item [(3)] if a string $\beta$ is enumerated in $T_1$ (or in $T_0$) above, 
then enumerate $\beta^c$ in $T_0$ (or in $T_1$, respectively).
\end{enumerate}

Clearly, \emph{$T_0$ and $T_1$ are recursively enumerable} because of this generating algorithm.
We observe that the sets $T_0$ and $T_1$ consist of 
\begin{itemize}
\item d-strings (11, 00, and those extending 10 enumerated at (2.i) and those extending 01 enumerated at (3) via~(2.i))
and 
\item p-strings (those extending $10$ enumerated at (2.ii) and those extending $01$ enumerated at (3) via (2.ii)).
\end{itemize}
We also observe that $11\in T_1$, $00\in T_0$, 
$T_0\cap T_1=\emptyset$, and $\alpha\in T_0 \Leftrightarrow \alpha^c\in T_1$.

\emph{Define a game $\omega[A]$ by $S\in \omega[A]$ if and only if $S$ has an initial segment in $T_1$}.
Lemma~\ref{k46} establishes computability of $\omega[A]$ 
(as well as the assertion that $T_0$ consists of losing determining strings and
 $T_1$ consists of winning determining strings)
by way of Proposition~\ref{delta0det2}.

\begin{lemma} \label{01incompatible}
Let $\alpha$, $\beta$ be distinct strings in $T_0\cup T_1$. Then $\alpha$ and $\beta$ are incompatible.
In particular, if $\alpha\in T_0$ and $\beta\in T_1$, then $\alpha$ and $\beta$ are incompatible.
\end{lemma}

\begin{proof}
Obviously, neither $\alpha$ nor $\beta$ is an empty string.
Since $T_0$ and $T_1$ consist of p-strings and d-strings,
there are three cases to consider:

\emph{Case}~(pp): \emph{Both $\alpha$ and $\beta$ are p-strings}.
Then either $\alpha$ or $\alpha^c$ is enumerated at (2.ii) of the generating algorithm 
and so $\alpha \in F$ or $\alpha^c\in F$.
Similarly, $\beta\in F$ or $\beta^c\in F$.
If  $\alpha \in F$ and $\beta\in F$,
then $\alpha$ and $\beta$ are incompatible, 
since any two distinct elements of $F$ are incompatible by Lemma~\ref{Fincompatible1}.
If $\alpha \in F$ and $\beta^c\in F$, then  $\alpha\supset 10$ and $\beta\supset 01$, 
so they are incompatible.
The other two subcases are similar.

\emph{Case}~(pd): \emph{one of $\alpha$ or $\beta$ is a p-string and the other is a d-string}.
Without a loss of generality, $\alpha$ is a p-string and $\beta$ is a d-string.
Suppose $\alpha$ and $\beta$ are compatible.  Then, $\beta\supset \alpha$.
In fact, $\beta^{--}\supseteq \alpha$.
As in (pp) above, either $\alpha \in F$ or $\alpha^c\in F$.
Also,  since either $\beta$ or $\beta^c$ is enumerated at (2.i) of the algorithm,
we have either 
(pd.i)~$\beta^{--}\subset \tilde{\beta}$ for some $\tilde{\beta}\in F$ or 
(pd.ii)~$(\beta^c)^{--}\subset \hat{\beta}$ for some $\hat{\beta}\in F$.
\emph{Subcase}: $\alpha\in F$ and (pd.i).   $\alpha$ and $\tilde{\beta}$ and both in $F$.  
So they are incompatible by Lemma~\ref{Fincompatible1}, contradicting the fact that $\alpha\subseteq \beta^{--}\subset \tilde{\beta}$.
\emph{Subcase}: $\alpha\in F$ and (pd.ii).  Then $\alpha\supseteq 10$ but $\beta\supset 01$, a contradiction.
\emph{Subcase}: $\alpha^c\in F$ and (pd.i).  Similar to the second subcase.
\emph{Subcase}: $\alpha^c\in F$ and (pd.ii).  Similar to the first subcase.

\emph{Case}~(dd): \emph{Both $\alpha$ and $\beta$ are d-strings}.
Immediate from Lemma~\ref{k41}.\end{proof}

\emph{Notation}.  We write $f(\beta)\!\downarrow \, \in A$ if $f(\beta)\in A$ (which requires $f(\beta)\downarrow$);
we write $f(\beta)\! \downarrow \, \notin A$ if $f(\beta)\downarrow$ but $f(\beta)\notin A$.

\begin{lemma}  \label{k43}
Let $\alpha\supset 1$ be a string of length $l_s$.
\begin{enumerate}
\item [\textup{(i)}] $\alpha$ extends a string in $T_1$ if and only if
 \textup{(i.a)} for some $s'\leq s$, $f(\alpha[l_{s'}]) \! \downarrow \, \in A$ (in this case, $\alpha[l_{s'}]\in T_1$) or 
\textup{(i.b)} $\alpha$ extends a d-string $\alpha'=(\alpha')^{--}*11$ 
such that  no prefix of $(\alpha')^{--}$ is in $F$ (in this case, $\alpha'\in T_1$).
 
 \item [\textup{(ii)}] $\alpha$ extends a string in $T_0$ if and only if 
\textup{(ii.a)} for some $s'\leq s$, $f(\alpha[l_{s'}]) \! \downarrow\, \notin A$ (in this case, $\alpha[l_{s'}]\in T_0$) or 
 \textup{(ii.b)} $\alpha$ extends a d-string $\alpha'=(\alpha')^{--}*00$ 
 such that no prefix of $(\alpha')^{--}$ is in $F$ (in this case, $\alpha'\in T_0$).
  
 \item [\textup{(iii)}] $\alpha$ does not extend a string in $T_0\cup T_1$ if and only if 
$\alpha$ is a p-string and no prefix of $\alpha$ is in~$F$.
\end{enumerate}
\end{lemma}
   
\begin{proof}
(i) ($\Longrightarrow$).
Assume $\alpha\supseteq 11$.  Then (i.b) is satisfied by letting $\alpha'=11$.

Assume $\alpha\supseteq 10$ extends a string $\alpha'\in T_1$. 
Suppose first that $\alpha'$ is enumerated in $T_1$ by applying (2.i) of the generating algorithm.  
(We show  (i.b) holds.)
Then $\alpha'=(\alpha')^{--}*11$ and $(\alpha')^{--}$ is properly extended by some element in~$F_s$.
Since any two different elements in $F$ are incompatible by Lemma~\ref{Fincompatible1}, 
no prefix of $(\alpha')^{--}$ is in $F$. 
So  (i.b) holds.
Suppose next that $\alpha'$ is enumerated in $T_1$ by applying (2.ii).
Then $f(\alpha')\in A$.
Since  $\alpha'=\alpha[l_{s'}]$ for some $s'\leq s$, we obtain (i.a).
Finally, the case where $\alpha'\supseteq 10$ is enumerated in $T_1$ by applying (3) is impossible,
since every string enumerated at (3) extends~$0$.

($\Longleftarrow$).
Assume $\alpha\supseteq 11$.  Since $11\in T_1$, the left hand side of (i) holds.

Assume $\alpha\supseteq 10$ and either (i.a) or (i.b) holds. 

Suppose (i.a) first.
By the definition of $f$,  $\alpha[l_{s'}]\in F_{s'}$.  
Since $f(\alpha[l_{s'}])\in A$, we have $\alpha[l_{s'}]\in T_1$ by (2.ii).
So $\alpha$ extends a string in $T_1$.

Suppose (i.b) next: $\alpha$ extends a d-string $\alpha'=(\alpha')^{--}*11$ 
such that no prefix of $(\alpha')^{--}$ is in $F$.  
We show that $\alpha'$ is in $T_1$.
 
Suppose $(\alpha')^{--}\subset \alpha[l_0]$ first. 
Since $l_0$ is even and $(\alpha')^{--}$ is a p-string of even length $<l_0$, we have $|(\alpha')^{--}|\leq l_0-2$. 
Since $l_0:=2k_0+2$, we can find a p-string $\beta$ of length $l_{0}$ that is an extension of $(\alpha')^{--}$ 
such that $\beta(2k_{0})=\varphi_{k_{0}}(2k_{0})$.  Then $\beta\in F_0$ and by (2.i) 
(for $\beta$ and $(\alpha')^{--}$ instead of $\alpha$ and $\alpha'$, respectively), 
$\alpha'=(\alpha')^{--}*11\in T_1$.

Otherwise, there is $s''$ such that $0<s''\leq s$ and
$\alpha[l_{s''-1}]\subseteq (\alpha')^{--}\subset \alpha[l_{s''}]$. 
Since $\alpha'$ is a d-string, $(\alpha')^{--}$ is a p-string. 
As $\alpha[l_{s''-1}]\subseteq (\alpha')^{--}$ and no prefix of $(\alpha')^{--}$ is in $F$, 
$\alpha[l_{s''-1}]$ is a p-string of which no prefix is in~$F$.
By Lemma~\ref{k42}, for each $t\leq s''-1$, we have $\alpha[l_{s''-1}](2k_t)=1-\varphi_{k_t}(2k_t)$.

Since $\alpha[l_{s''-1}]\subseteq (\alpha')^{--}\subset \alpha[l_{s''}]$,  we have $l_{s''-1}<l_{s''}$.
Hence $l_{s''}:=\max\{l_{s''-1}, 2k_{s''}+2\}=2k_{s''}+2$.
Since $| (\alpha')^{--}|$ and $l_{s''}$ are even, $|(\alpha')^{--}| \le 2k_{s''}$. 
We can find a p-string $\beta$ of length $l_{s''}$ that is an extension of $(\alpha')^{--}$ 
such that $\beta(2k_{s''})=\varphi_{k_{s''}}(2k_{s''})$. 
Therefore, for each $t\leq s''-1$, 
we have $\beta[l_{s''-1}](2k_t)=(\alpha')^{--}[l_{s''-1}](2k_t)=1-\varphi_{k_t}(2k_t)$.
So $\beta\in F_{s''}$ by (1). 
Then since $|(\alpha')^{--}|\ge l_{s''-1}$, we have by (2.i) (for $\beta$ and $(\alpha')^{--}$ instead of $\alpha$ and $\alpha'$, respectively), $\alpha'=(\alpha')^{--}*11\in T_1$.

(ii)~Similar to~(i). 

(iii) ($\Longrightarrow$).
Suppose that $\alpha$ does not extend a string in $T_0\cup T_1$.  
Then the negations of (i.a) and of (ii.a) imply for each $t\leq s$, $f(\alpha[l_{t}])\uparrow$, 
which implies by Lemma~\ref{k42} that no prefix of $\alpha$ is in $F$. 
Furthermore, (since no prefix of $\alpha$ is in $F$) the negations of (i.b) and of (ii.b) imply
that $\alpha$ does not extend a d-string. By Lemma~\ref{k41} (i), $\alpha$ is a p-string.

 ($\Longleftarrow$).
Suppose that $\alpha$ is a p-string and no prefix of $\alpha$ is in $F$. 
Since $\alpha$ is a p-string, no prefix of $\alpha$ is a d-string. So $\alpha$ does not satisfy (i.b) or (ii.b). 
Since no prefix $\alpha'$ of $\alpha$ is in~$F$, we have for such $\alpha'$, $f(\alpha')\uparrow$. 
So $\alpha$ does not satisfy (i.a) or (ii.a). 
Therefore, $\alpha$ does not extend a string in $T_0\cup T_1$.\end{proof}

\begin{lemma} \label{k45}
Let $\alpha\supset 1$ be a string of length $l_s$ such that $\alpha(2k_s)=\varphi_{k_s}(2k_s)$.
Then $\alpha$ extends a  string in $T_0\cup T_1$.
\end{lemma}

\begin{proof}
If $\alpha\supseteq 11$, the conclusion follows immediately, since $11\in T_1$.

Suppose $\alpha\supseteq 10$.  We prove the lemma by induction on~$s$.
Assume $s=0$. If $\alpha$ is a p-string, then  $\alpha\in F_0$.
By (2.ii) of the generating algorithm for $T_0$ and $T_1$, we obtain $\alpha\in T_0\cup T_1$.
Otherwise, by Lemma \ref{k41} (i), $\alpha$ extends a d-string $\beta$. 
Since $|\beta^{--}|<l_0\le l_s$ for all~$s$,  no prefix of $\beta^{--}$ is in~$F$
 (because $F$ consists of certain strings of length $l_s$ for some~$s$).
 By Lemma \ref{k43} (i.b) or (ii.b),  $\alpha$ extends a string (namely $\beta$) in $T_0\cup T_1$.

Assume the lemma holds for $s-1$. If for some $s'<s$, $\alpha(2k_{s'})=\varphi_{k_{s'}}(2k_{s'})$
then by the induction hypothesis, $\alpha[l_{s'}]$ extends a  string in $T_0\cup T_1$. 
So $\alpha$  extends a string in $T_0\cup T_1$.  
Otherwise,  for each $s'<s$,  $\alpha(2k_{s'})=1-\varphi_{k_{s'}}(2k_{s'})$. 
If $\alpha$ is a p-string then $\alpha\in F$ by (1), hence it is in $T_0\cup T_1$ by (2.ii) of the construction.
If $\alpha$ is not a p-string then  by Lemma \ref{k41} (i), $\alpha$ extends a d-string $\beta$. 
Then $|\beta^{--}|<l_s$. Since $\beta\subseteq\alpha$ and 
for each $s'<s$,  $\alpha(2k_{s'})=1-\varphi_{k_{s'}}(2k_{s'})$, no prefix of $\beta^{--}$ is in $F$ by~(1). 
By Lemma \ref{k43} (i.b) or (ii.b), $\alpha$ extends a string (namely $\beta$) in $T_0\cup T_1$.%
\end{proof}

\begin{lemma} \label{k46} 
Any coalition $S\in\REC$ has an initial segment in $T_0$ or in $T_1$, but not both.
\end{lemma}

\begin{proof}
We show that $S$~has an initial segment in $T_0\cup T_1$. 
Lemma~\ref{01incompatible} implies that
 $S$ does not have initial segments in both $T_0$ and $T_1$.
(We can actually show that $S$ has exactly one initial segment in $T_0\cup T_1$,
a fact used to construct a type~4 game in Section~\ref{examples:nocarrier}.)

If $S\supseteq 1$, suppose $\varphi_k$ is the characteristic function for~$S$. 
Then $k\in\{k_0,k_1,k_2, \ldots\}$ since this set contains the set $\CRec$ of characteristic indices.
So $k=k_s$ for some~$s$.
By Lemma~\ref{k45}, the initial segment $S[l_s]$ (i.e., $\varphi_{k_s}[l_s]$) extends
a string in $T_0\cup T_1$.  So, $S$ has an initial segment in $T_0\cup T_1$.

If $S\supseteq 0$, then $S^c\supseteq 1$ has an initial segment in $T_0\cup T_1$
by the argument above.  So, $S$ has an initial segment in $T_1\cup T_0$.\end{proof}

\bigskip

Next, we show that the game~$\omega[A]$ has the desired properties.  Before showing monotonicity, 
we need the following lemma.  For strings $\alpha$ and $\beta$ with $|\alpha|\le |\beta|$, 
we say \emph{$\beta$ properly contains~$\alpha$}
if for each $k<|\alpha|$, $\alpha(k)\leq\beta(k)$ and for some $k'<|\alpha|$,  $\alpha(k')<\beta(k')$; 
we say \emph{$\beta$~is properly contained by~$\alpha$}
 if for each $k<|\alpha|$, $\beta(k)\le \alpha(k)$ and for some $k'<|\alpha|$, $\beta(k')<\alpha(k')$.

\begin{lemma} \label{k44}
Let $\alpha$ and $\beta$ be strings such that $l_s=|\alpha|\le |\beta|$ for some~$s$.
\textup{(i)}~If $\alpha$ extends a string in $T_1$ and $\beta$ properly contains~$\alpha$, 
then $\beta$ extends a string in~$T_1$.
\textup{(ii)}~If $\alpha$ extends a string in $T_0$ and $\beta$ is properly contained by~$\alpha$, 
then $\beta$ extends a string in~$T_0$.
\end{lemma}

\begin{proof}
We only prove~(i).  The proof for (ii) is similar.
Suppose that $\alpha$ extends a string in $T_1$ and that $\beta$ properly contains~$\alpha$.

\emph{Case}~1: $\alpha\supseteq 1$. 
In this case, (i.a) or (i.b) of Lemma~\ref{k43} holds.

\emph{First assume \textup{(i.a)} is the case}: we can choose an $s'\leq s$ such that
$f(\alpha[l_{s'}])\!\!\downarrow \in A$ (in this case, $\alpha[l_{s'}]\in T_1$).
If $\beta$ extends $\alpha[l_{s'}]$, clearly the conclusion  holds. 
Otherwise, since $|\beta|\geq l_s\geq l_{s'}$, $\alpha[l_{s'}]$ and $\beta$ are incompatible; that is,
there exists $k< l_{s'}$ such that $\alpha[l_{s'}](k)\neq \beta(k)$.  Choose the least such $k$; since 
$\beta$ properly contains~$\alpha$, we have $\alpha[l_{s'}](k)=0$ and $\beta(k)=1$.
Let $\beta'=\beta[k] (=\alpha[k])$.
Note that $f(\alpha[l_{s'}])\downarrow$ implies $\alpha[l_{s'}]\in F$, which in turn implies
$\alpha[l_{s'}]$ is a p-string. 

Suppose $k$ is even.  We will show that $\beta$ extends $\beta'*11\in T_1$.
Since $k< l_{s'}$ and $l_{s'}$ is also even, we have $k+1<l_{s'}$, so that $\alpha[l_{s'}](k+1)\downarrow$.
Since $\alpha[l_{s'}]$ is a p-string,
$\beta(k+1)\geq\alpha[l_{s'}](k+1)=1-\alpha[l_{s'}](k)=1$. 
So $\beta(k) \beta(k+1)=11$. 
Hence $\beta'*11\subseteq\beta[l_s]$. 
Since $\alpha[l_{s'}]\in F$,  no proper prefix of $\alpha[l_{s'}]$ is in~$F$. 
As $\beta'\subset \alpha[l_{s'}]$, no prefix of $\beta'$ is in~$F$. 
So by Lemma \ref{k43} (i.b), $\beta[l_s]$ extends a string (namely, $\beta'*11$) in~$T_1$.

Suppose $k$ is odd.  We will show that $\beta$ extends $(\beta')^-*11\in T_1$.
Since $\alpha[l_{s'}]$ is a p-string, $\beta(k-1)=\alpha[l_{s'}](k-1)=1-\alpha[l_{s'}](k)=1$. 
So $\beta(k-1) \beta(k)=11$. 
Hence $(\beta')^-*11\subseteq\beta[l_s]$. 
Since no proper prefix of $\alpha[l_{s'}]$ is in~$F$ and $(\beta')^- \subset \alpha[l_{s'}]$,
no prefix of $(\beta')^-$ is in~$F$. 
So by Lemma \ref{k43} (i.b), $\beta[l_s]$ extends a string (namely, $(\beta')^-*11$) in~$T_1$.
 
\emph{Next assume \textup{(i.b)} is the case}: $\alpha$ extends a d-string $\alpha'=(\alpha')^{--}*11$ 
such that no prefix of $(\alpha')^{--}$ is in~$F$ (in this case, $\alpha'\in T_1$).
Choose the least $k\le |\alpha|$ such that $\alpha(k)\neq \beta(k)$; we have $\alpha(k)=0$ and $\beta(k)=1$.
Let $\beta'=\beta[k] (=\alpha[k])$.
Since $\alpha'(|\alpha'|-2)=\alpha'(|\alpha'|-1)=1$, either $k>|\alpha'|-1$ or $k<|\alpha'|-2=|(\alpha')^{--}|$.
If $k>|\alpha'|-1$, we get $\beta'\supseteq \alpha'$.
This implies $\beta\supseteq \beta'\supseteq \alpha'\in T_1$;
hence $\beta$ extends a string in~$T_1$.  
Otherwise, we have $k<l:=|(\alpha')^{--}|$ and $\beta'\subset (\alpha')^{--}$.

Suppose $k$ is even.
Since $k<l$ and $l$ is also even, we have $k+1<l$, so that $(\alpha')^{--}(k+1)\downarrow$.
Since $\alpha$ is a p-string, $\beta(k+1)\geq (\alpha')^{--}(k+1)=1-(\alpha')^{--}(k)=1$. 
So $\beta(k) \beta(k+1)=11$. 
Hence $\beta'*11\subseteq \beta[l_s]$. 
Since no prefix of $(\alpha')^{--}$ is in~$F$ and $\beta'\subset (\alpha')^{--}$,  
no prefix of $\beta'$ is in~$F$. 
So by Lemma \ref{k43} (i.b), $\beta[l_s]$ extends a string (namely, $\beta'*11$) in~$T_1$.

Suppose $k$ is odd.
Since $(\alpha')^{--}$ is a p-string, $\beta(k-1)=(\alpha')^{--}(k-1)=1-(\alpha')^{--}(k)=1$. 
So $\beta(k-1) \beta(k)=11$.
Hence $(\beta')^-*11\leq\beta[l_s]$. 
Since no prefix of $(\alpha')^{--}$ is in~$F$ and $(\beta')^-\subset (\alpha')^{--}$,
no prefix of $(\beta')^-$ is in $F$. 
So by Lemma \ref{k43} (i.b), $\beta[l_s]$ extends a string (namely, $(\beta')^-*11$) in~$T_1$.

\emph{Case}~2: $\alpha\supseteq 0$. 
First note that assertion (ii) for Case~1 can be proved by an argument similar to 
the proof of assertion (i) for Case~1 above (use Lemma \ref{k43} (ii) instead of Lemma \ref{k43} (i)).
By the construction of $T_1$ and $T_0$, 
$\alpha^c\supseteq 1$ extends a string in $T_0$ and $\beta^c$ is properly contained by~$\alpha^c$.
Applying assertion (ii) for Case~1, we obtain that $\beta^c$ extends a string in~$T_0$.
Hence $\beta$ extends a string in~$T_1$.\end{proof}

Note that the preceding proof shows that $\beta$ actually extends a \emph{d-string} 
unless it extends $\alpha[l_{s'}]$.

\begin{lemma} \label{k47a}
The game $\omega[A]$ is monotonic.
\end{lemma}

\begin{proof}
Suppose $B\in \omega[A]$ and $B'\supseteq B$.  
By the definition of $\omega[A]$, $B$ has an initial segment $\alpha\in T_1$.
Choose the least $s$ such that $l_s\ge |\alpha|$.
Then the initial segment $B[l_s]$ extends $\alpha\in T_1$.
Let $\beta=B'[l_s]$.
Then either $\beta=B[l_s]$ or $\beta$ properly contains $B[l_s]$.

If $\beta=B[l_s]$, then clearly $\beta$ extends $\alpha\in T_1$ and so does~$B'$.
Therefore, $B'\in \omega[A]$.
Otherwise, $\beta$ properly contains $B[l_s]$, which extends $\alpha\in T_1$.
By Lemma~\ref{k44} (i), $\beta$ extends a string in $T_1$ and so does~$B'$.
Therefore, $B'\in \omega[A]$.\end{proof}

\begin{lemma} \label{k47b}
The game $\omega[A]$ is proper and strong.
\end{lemma}

\begin{proof}
It suffices to show that $S^c\in\omega \Leftrightarrow S\notin\omega$.
From the observations that
$T_0$ and $T_1$ consist of determining strings and
that $\alpha^c\in T_0 \Leftrightarrow \alpha \in T_1$, we have:
$S^c \in\omega$  iff  $S^c$ has an initial segment in~$T_1$
						 iff  $S$ has an initial segment in~$T_0$
						 iff  $S\notin\omega$.\end{proof}
 
\begin{lemma} \label{k48} 
The game $\omega[A]$ is nonweak and does not have a finite carrier.
\end{lemma}

\begin{proof}
We construct a set~$B$ such that for infinitely many~$l$,
the $l$-initial segment $B[l]$ has an extension that is winning and an extension that is losing.  
Let $B\supseteq 10$ be a set such that for each $k_s$, $B(2k_s)=1-\varphi_{k_s}(2k_s)$ and
any initial segment of $B$ of even length is a p-string. 
Let $s$ be such that $l_{s+1}>l_s$. 

Then $l_{s+1}:=\max\{l_s,2k_{s+1}+2\}=2k_{s+1}+2$ and 
$2k_{s+1}+2>l_s$ implies (since both sides are even numbers) that $2k_{s+1}\geq l_s$.
By the definition of $B$, for each $t\leq s$, we have $B(2k_{t})=1-\varphi_{k_{t}}(2k_{t})$
and $2k_t<l_{s}$ 
(the last inequality from the observation that $l_s>2k_s+1$, $2k_{s-1}+1$, $2k_{s-2}+1$, \ldots, $2k_0-1$). 
Then since $2k_{s+1}\geq l_{s}$, there is a p-string  $\alpha\supseteq B[l_{s}]$ of length $l_{s+1}$ such that 
$\alpha(2k_{s+1})=\varphi_{k_{s+1}}(2k_{s+1})$ and for each $t\leq s$, $\alpha(2k_{t})=1-\varphi_{k_{t}}(2k_{t})$. 
Then by (1), $\alpha\in F_{s+1}$ and $|\alpha^{--}|=|\alpha|-2=l_{s+1}-2= 2k_{s+1}\geq l_{s}$. 
So by (2.i) of the generating algorithm, $\alpha^{--}*11\in T_1$ and $\alpha^{--}*00\in T_0$.

There are infinitely many such $s$. It follows that 
any initial segment of $B$ has an extension in $T_1$ and an extension in~$T_0$. 
This means that the game has no finite carrier.

To show nonweakness, we give three (winning) coalitions in $T_1$ whose intersection is empty.
First, $10$ (in fact any initial segment of the coalition $B \supseteq 10$)
has extensions $\alpha$ in $T_1$ and $\beta$ in~$T_0$ by the argument above.
So $01$ has the extension~$\beta^c$ in~$T_1$.
Clearly, the intersection of the winning coalitions $11\in  T_1$, $\alpha\supseteq 10$, and 
$\beta^c\supseteq 01$ is empty.\end{proof}

Note that the proof that $\omega[A]$ has no finite carrier depends on (2.i), but not (2.ii) or (3),
of the generating algorithm.

\subsection{Infinite computable games}
\label{examples:nocarrier}

In this section,
for each of the ten conventional types not shown to be empty so far (footnote~\ref{emptytypes}),
we give an example of an infinite computable game of that type.
Most examples are based on the game $\omega[A]$ in Section~\ref{nice_games}.

\begin{enumerate}
\item[1] $(++++)$
A monotonic, proper, strong,  nonweak game.  $\omega[A]$ is such a game. 

\item[3] {$(++-+)$} A monotonic, proper, nonstrong, nonweak game.
Let $\omega=\omega[\emptyset]\cap\omega[\N]$; that is, 
$S\in \omega$ if and only if $S\in \omega[\emptyset]$ and $S\in \omega[\N]$.

To show $\omega$ is proper, suppose $S\in \omega$ and $S^c\in \omega$.
Then $S\in \omega[\N]$ and $S^c\in \omega[\N]$, contradicting the properness of~$\omega[\N]$.

To show $\omega$ is nonstrong, let $\alpha\in F$.  We show that both $\alpha$ and $\alpha^c$ are losing.
On the one hand, we have $\alpha\in T_0^\emptyset$ by (2.ii) of the generating algorithm.
Since $T_0^\emptyset$ consists of losing determining strings, $\alpha\notin \omega[\emptyset]$.
Hence $\alpha\notin \omega$.
On the other hand, we have $\alpha\in T_1^\N$ by (2.ii).  Hence $\alpha^c\in T_0^\N$.
Since $T_0^\N$ consists of losing determining strings, $\alpha^c\notin \omega[\N]$.
Hence $\alpha^c\notin \omega$, as desired.

Computability, monotonicity, and nonweakness of~$\omega$ are immediate from the corresponding properties
of~$\omega[A]$.
The proof that $\omega$ does not have a finite carrier is similar to the proof for~$\omega[A]$. 

\item[4] {$(++--)$} A monotonic, proper, nonstrong, weak game.
In the construction of (the sets $T_0$ and $T_1$ for) $\omega[A]$ in Section \ref{nice_games}, 
replace (2.i), (2.ii), and (3) by 
\begin{enumerate}
\item [(2*.i)] for each p-string $\alpha'$ that is a proper prefix of
$\alpha$, if $s=0$ or $|\alpha'|\geq l_{s-1}$, then enumerate
$1*\alpha'*11$ in~$T_1$ and $1*\alpha'*00$ in~$T_0$;
furthermore, enumerate $0$ in~$T_0$;
\item [(2*.ii)] if $f(\alpha)\in A$, enumerate $1*\alpha$ in $T_1$; 
if $f(\alpha)\notin A$, enumerate $1*\alpha$ in $T_0$;
\item [(3*)] if a string $\beta=1*\beta'$ is enumerated in $T_1$ (or in
$T_0$) above, 
then enumerate $1*(\beta')^c$ in $T_0$ (or in $T_1$, respectively).
\end{enumerate}
Let $T'_0$ and $T'_1$ be the sets $T_0$ and $T_1$ in the original (Section \ref{nice_games}) 
construction of $\omega[A]$ renamed.
We observe that $\beta=1*\beta'\in T_i$ if and only if $\beta'\in T'_i$.

We first show that any coalition $S$ has exactly one initial segment in $T_0\cup T_1$.
This is immediate if $S\supseteq 0$.
So, suppose $S\supseteq 1$.
Define $S'$ by $S'(k)=S(k+1)$ for all $k$.
Then, by the proof of Lemma~\ref{k46} for $\omega[A]$, 
$S'$ has exactly one initial segment $S'[k]$ in $T'_0\cup T'_1$.
From the observation above, $S[k+1]=1*S'[k]\in T_0\cup T_1$ for a unique~$k$, which is what we wanted.

To show the game is monotonic, it suffices to show Lemma~\ref{k44} (i) holds for the newly defined game.
Suppose that $\alpha$, $\beta$ satisfy the assumption of the lemma and that
$\alpha$ extends a string $\hat{\alpha}$ in $T_1$ and $\beta$ properly contains~$\alpha$.
Then, $\hat{\alpha}\supseteq 1$; write $\hat{\alpha}=1*\hat{\alpha}'$.
Then $\hat{\alpha}'\in T'_1$ from the observation above.
We can write $\beta=1*\beta'$.
Then $\beta'$ either extends or properly contains $\hat{\alpha}'\in T'_1$.
If $\beta'$ extends $\hat{\alpha}'\in T'_1$, then $\beta$ extends $1*\hat{\alpha}'\in T_1$, as desired.
Otherwise, $\beta'$ properly contains $\hat{\alpha}'\in T'_1$.
By Lemma~\ref{k44} for the original game $\omega[A]$
(the condition that $l_s=|\alpha|$ can be ignored for our purpose), 
$\beta'$ extends a string $\hat{\beta}\in T'_1$.
So, $\beta=1*\beta'$ extends $1*\hat{\beta}\in T_1$, as desired.

The game is weak (hence proper by Lemma~\ref{weakisproper})
since every winning coalition extends~$1$; in other words, $0$ is a veto player.
It is nonstrong since $\{0\}\supseteq 100\in T_0$ implies $\{0\}\notin\omega$,
 while $\{0\}^c \supseteq 0\in T_0$ implies $\{0\}^c \notin\omega$.
The proof that the game is computable and has no finite carrier is
 similar to the proofs for~$\omega[A]$.

\item[5] {$(+-++)$} A monotonic, nonproper, strong, nonweak game.
Let $\omega=\omega[\emptyset]\cup\omega[\N]$; that is,  
$S\in \omega$ if and only if $S\in \omega[\emptyset]$ or $S\in \omega[\N]$.

To show $\omega$ is nonproper, let $\alpha\in F$.  We show that both $\alpha$ and $\alpha^c$ are winning.
On the one hand, we have $\alpha\in T_1^\N$ by (2.ii).  So $\alpha\in \omega[\N]$, implying $\alpha\in \omega$.
On the other hand, we have $\alpha\in T_0^\emptyset$ by (2.ii).  Hence $\alpha^c\in T_1^\emptyset$.
So $\alpha^c\in \omega[\emptyset]$.
Hence $\alpha^c\in \omega$, as desired.

To show $\omega$ is strong, suppose $S\notin \omega$ and $S^c\notin \omega$.
Then $S\notin \omega[\N]$ and $S^c\notin \omega[\N]$, contradicting the strongness of~$\omega[\N]$.

Computability and monotonicity of~$\omega$ are immediate from the corresponding properties
of~$\omega[A]$.  Nonweakness is immediate from nonproperness by Lemma~\ref{weakisproper}.
The proof that $\omega$ does not have a finite carrier is similar to the proof for~$\omega[A]$. 

\item[7] {$(+--+)$} A monotonic, nonproper, nonstrong, nonweak game.
Let $A$ be the set of even numbers. 
In the construction of $\omega[A]$, replace (2.ii) and (3) by 
\begin{enumerate}
\item [(2*.ii)] if $f(\alpha)\in A$, enumerate $\alpha$ and $\alpha^c$ in $T_1$; 
if $f(\alpha)\notin A$, enumerate $\alpha$ and $\alpha^c$ in $T_0$;
\item [(3*)] if a string $\beta$ is enumerated in $T_1$ (or in $T_0$) by applying (2.i), 
then enumerate $\beta^c$ in $T_0$ (or in $T_1$, respectively).
\end{enumerate}

To show the game is monotonic, it suffices to show Lemma~\ref{k44} (i) holds for the newly defined game.
Suppose that $\alpha$, $\beta$ satisfy the assumption of the lemma and that
$\alpha$ extends a string $\alpha'$ in $T_1$ and $\beta$ properly contains~$\alpha$.
Let $T'_0$ and $T'_1$ be the sets $T_0$ and $T_1$ in the original 
construction of $\omega[A]$ renamed.
Note that the replacement of (2.ii) and (3) by (2*.ii) and (3*) only affects p-strings, but not d-strings;
hence the set of d-strings in $T_1$ is the same as the set of d-strings in $T'_1$,
the set of d-strings in $T_0$ is the same as the set of d-strings in $T'_0$, and
the set of p-strings in $T_0\cup T_1$ is the same as the set of p-strings in $T'_0\cup T'_1$.
If $\alpha'$ is a d-string in $T_1$, it is in $T'_1$.
Lemma~\ref{k44} (i) (for the original game) implies that $\beta$ extends a string in~$T'_1$.
In fact, an inspection of the proof of Lemma~\ref{k44} reveals that $\beta$ extends a d-string in~$T'_1$,
unless $\beta\supseteq \alpha'$, in which case the conclusion is obvious.  So assume $\beta\not\supseteq \alpha'$.
Then $\beta$ extends a d-string in~$T'_1$; hence it extends a d-string in~$T_1$, as desired.
If $\alpha'$ is a p-string in $T_1$, it is in $T'_1\cup T'_0$.
If $\alpha'\in T'_1$, then Lemma~\ref{k44} (i) implies that $\beta$ extends a string in~$T'_1$.
So the rest of the proof is similar.
If $\alpha'\in T'_0$, then Lemma~\ref{k44} (ii) implies that $\beta^c$ extends a string in~$T'_0$.
Assume $\beta\not\supseteq \alpha'$ as before.  
Then $\beta^c$ extends a d-string in~$T'_0$; hence it extends a d-string in~$T_0$.
By (3*), $\beta$ extends a d-string in~$T_1$, as desired.

The game is nonproper since (2*.ii) implies that there is a string $\alpha\in F$ such that
the coalitions $\{i: \alpha(i)=1\}$ and $\{i: \alpha(i)=1\}^c$
(which extends $\alpha^c$) are winning.
Similarly, it is nonstrong since there is a string $\alpha\in F$ such that
the coalitions above are losing.
It is nonweak by Lemma~\ref{weakisproper} since it is nonproper.
The proof that the game is computable and has no finite carrier is
 similar to the proofs for~$\omega[A]$.

\item[9] {$(-+++)$} A nonmonotonic, proper, strong, nonweak game.
In the construction of $\omega[A]$, replace (2.i) by 
\begin{enumerate}
\item [(2*.i)] for each p-string $\alpha'\neq \emptyset$ that is a proper prefix of $\alpha$, 
	if $s=0$ or $|\alpha'|\geq l_{s-1}$, then enumerate $\alpha'*11$ in $T_1$ and $\alpha'*00$ in~$T_0$;
	furthermore, enumerate $00$ in~$T_1$.
\end{enumerate}
By (3) of the construction, $11\in T_0$.  
(In other words, the game is constructed from the sets $T_0:=T'_0\cup \{11\}\setminus\{00\}$  
and $T_1:=T'_1\cup \{00\}\setminus \{11\}$, 
where $T'_0$ and $T'_1$ are $T_0$ and $T_1$ in the original construction of $\omega[A]$ renamed.)
Since $00$ is winning and $11$ is losing, 
the game is nonmonotonic.  It is also nonweak since $00$ (or an empty coalition) is winning.
For the remaining properties, the proofs  are similar to the proofs for~$\omega[A]$. 

\item[11] {$(-+-+)$} A nonmonotonic, proper, nonstrong, nonweak game.
In the construction of $\omega[A]$, replace (2.i) and (3) by 
\begin{enumerate}
\item [(2*.i)] for each p-string $\alpha'\neq \emptyset$ that is a proper prefix of~$\alpha$, 
	if $s=0$ or $|\alpha'|\geq l_{s-1}$, then enumerate $\alpha'*11$ in $T_1$ and $\alpha'*00$ in~$T_0$;
	furthermore, enumerate $00$ and $11$ in~$T_0$;
\item [(3*)] if a string $\beta\notin\{00,11\}$ is enumerated in $T_1$ (or in $T_0$) above, 
then enumerate $\beta^c$ in $T_0$ (or in $T_1$, respectively).
\end{enumerate}
(In other words, the game is constructed from the sets $T_0:=T'_0\cup \{11\}$  
and $T_1:=T'_1\setminus \{11\}$, 
where $T'_0$ and $T'_1$ are $T_0$ and $T_1$ in the original construction of $\omega[A]$ renamed.)

The game is nonmonotonic since $N$ is losing but there are winning coalitions.
It is proper since it is a subset of $\omega[A]$, which is proper.
It is nonstrong since $11$, $00\in T_0$ implies that the coalitions $\{0, 1\}$, $\{0,1\}^c$ are losing.

To show nonweakness, find a $\beta\in T_1$ such that $|\beta|=l_{t+1}$ for some~$t$
(e.g., let $\beta=\alpha^{--}*11$ in the proof of Lemma~\ref{k48}, 
with $s$ replaced by $t$).
Choose an $s$ such that $l_{t+1}<l_s<l_{s+1}$.
Following the proof of Lemma~\ref{k48}, 
we can find $\alpha\in F_{s+1}$ such that
$|\alpha^{--}|\ge l_s$, $\alpha^{--}*11\in T_1$, and $\alpha^{--}*00 \in T_0$.
Then $(\alpha^c)^{--}*11\in T_1$.
Nonweakness follows since the intersection of winning coalitions 
$\beta$ (regarded as the coalition $\{i: \beta(i)=1\}$),
$\alpha^{--}*11\in T_1$, and $(\alpha^c)^{--}*11$ is empty.

The proofs of computability and nonexistence of a finite carrier are similar to the proofs for~$\omega[A]$.

\item[12] {$(-+--)$} A nonmonotonic, proper, nonstrong, weak game.
Let $A=\N$.
In the construction of $\omega[A]=\omega[\N]$, replace (2.i) by 
\begin{enumerate}
\item [(2*.i)] for each p-string $\alpha'$ that extends $1010$ or $1001$ and is a proper prefix of~$\alpha$, 
	if $s=0$ or $|\alpha'|\geq l_{s-1}$, then enumerate $\alpha'*11$ in $T_1$ and $\alpha'*00$ in~$T_0$;
	furthermore, enumerate d-strings $11$ and $1000$ in~$T_1$
	and strings $1011$ and $0$ in~$T_0$.
\end{enumerate}
and remove~(3).  
To show that any coalition $S$ has an initial segment in $T_0\cup T_1$, suppose that
$S$~extends $1010$ or $1001$.  (The other cases are immediate.)
Let $T'_0$ and $T'_1$ be $T_0$ and $T_1$ in the original construction of $\omega[\N]$ renamed.
Then, by Proposition~\ref{k46}, $S$ has an initial segment $S[k]$ in $T'_0\cup T'_1$,
where $k\geq 4$ without loss of generality.
If $S[k]$ is enumerated in $T'_0\cup T'_1$ by applying (2.ii), then it is enumerated 
in $T_0\cup T_1$ by applying (2.ii).  So, the conclusion follows.
If $S[k]$ is enumerated in $T'_0\cup T'_1$ by applying (2.i), then 
$S[k]$ is equal to $\alpha'*11$ or $\alpha'*00$ for some p-string $\alpha'$ satisfying the requirements in (2.i).
Clearly, $\alpha'$ extends $1010$ or $1001$.  So, $S[k]$ is enumerated in 
$T_0\cup T_1$ by applying (2*.i).  So the conclusion follows.

To show that no coalition $S$ has initial segments in both $T_0$ and $T_1$,
it suffices to show that a string $\alpha$ enumerated in $T_0$ by (2*.i) and a p-string $\beta$ enumerated in $T_1$
by (2.ii) are incompatible.  (Note that all $\alpha\in F$ are enumerated in $T_1$ and none in $T_0$ by (2.ii).)
Since $\beta\supset 10$, it is incompatible with $0\in T_1$.  All the other strings enumerated by (2*.i) are
d-strings, so $\alpha$ and $\beta$ are compatible only if $\alpha$ extends $\beta$, 
which in turn extends (since $\beta\in F$ is of length $\geq 4$)
 $1001$ or $1010$.  Then, $\alpha=\alpha'*00$ for some $\alpha'$, so as above, $\alpha\in T'_0$;
 similarly, $\beta\in T'_1$.  This implies that $\alpha$ and $\beta$ are incompatible.

The game $\omega$ defined above is nonmonotonic since $1000$ is winning but $1011$ is not.
To see $\omega$ is weak (hence proper by Lemma~\ref{weakisproper}), note that any winning
coalition extends $1$; so the intersection contains a veto player~$0$.
The game is nonstrong because $0$, $1011\in T_0$ imply that the coalitions $\{1\}$ and $\{1\}^c$ are losing.
The proofs of computability and nonexistence of a finite carrier are similar to the proofs for~$\omega[A]$.

\item[13] {$(--++)$} A nonmonotonic, nonproper, strong, nonweak game.
In the construction of $\omega[A]$, replace (2.i) and (3) by
\begin{enumerate}
\item [(2*.i)] for each p-string $\alpha'\neq \emptyset$ that is a proper prefix of~$\alpha$, 
	if $s=0$ or $|\alpha'|\geq l_{s-1}$, then enumerate $\alpha'*11$ in $T_1$ and $\alpha'*00$ in~$T_0$;
	furthermore, enumerate $00$ and $11$ in~$T_1$;
	\item [(3*)] if a string $\beta\notin\{00,11\}$ is enumerated in $T_1$ (or in $T_0$) above, 
then enumerate $\beta^c$ in $T_0$ (or in $T_1$, respectively).
\end{enumerate}
(In other words, the game is constructed from the sets $T_0:=T'_0\setminus \{00\}$  
and $T_1:=T'_1\cup \{00\}$, 
where $T'_0$ and $T'_1$ are $T_0$ and $T_1$ in the original construction of $\omega[A]$ renamed.)

The game is nonmonotonic since $\emptyset$ is winning but there are losing coalitions.
It is nonproper since the coalitions $\{0, 1\}$, $\{0,1\}^c$ are winning.
It is strong since its subset $\omega[A]$ is strong.
It is nonweak by Lemma~\ref{weakisproper} since it is nonproper.
The proofs of computability and nonexistence of a finite carrier are similar to the proofs for~$\omega[A]$.

\item[15] {$(---+)$} A nonmonotonic, nonproper, nonstrong, nonweak game.
In the construction of $\omega[A]$, replace (2.i) and (3) by 
\begin{enumerate}
\item [(2*.i)] for each p-string $\alpha'$ that extends $1010$ or $1001$ and is a proper prefix of~$\alpha$, 
	if $s=0$ or $|\alpha'|\geq l_{s-1}$, then enumerate $\alpha'*11$ in $T_1$ and $\alpha'*00$ in~$T_0$;
	furthermore, enumerate d-strings $00$, $1000$, and $0111$ in~$T_0$
	and d-strings $11$, $1011$ and $0100$ in~$T_1$;
\item [(3*)] if a string $\beta\notin\{00, 11, 1000, 0111, 1011, 0100\}$ is enumerated in $T_1$ 
(or in $T_0$) above, 
then enumerate $\beta^c$ in $T_0$ (or in $T_1$, respectively).
\end{enumerate}
The game is nonmonotonic since  $0100$ is winning but $0111$ is not.
The game is nonproper since $1011$, $0100\in T_1$ imply that the coalitions $\{1\}$ and $\{1\}^c$ are winning.
It is nonstrong since $1000$, $0111\in T_0$  imply $\{0\}$ and $\{0\}^c$ are losing.  
It is nonweak by Lemma~\ref{weakisproper} since it is nonproper. 
The proofs of computability and nonexistence of a finite carrier are similar to the proofs for~$\omega[A]$.

\end{enumerate}






\end{document}